\documentclass[11pt,onecolumn] {IEEEtran}
\usepackage{amsmath}
\usepackage{times}
\usepackage{amsbsy,color}
\usepackage{latexsym}
\usepackage{amssymb}
\usepackage{graphicx}
\usepackage{subfigure}
\usepackage{amsfonts}
\usepackage{epsfig,subfigure,tikz}
\begin{document}

\title{\huge Study of Distributed Conjugate Gradient Strategies for Distributed Estimation Over Sensor Networks }
\author{Songcen~Xu*,~
        Rodrigo C. de~Lamare,~\IEEEmembership{Senior Member,~IEEE,}
        and~H. Vincent~Poor,~\IEEEmembership{Fellow,~IEEE}  \vspace{-1em}
\thanks{S. Xu* is with the Communications and Signal Processing Research Group, Department of Electronics, University of York, YO10 5DD York, U.K.
(e-mail: songcen.xu@york.ac.uk).}
\thanks{R. C. de Lamare is with CETUC / PUC-Rio, Brazil and Department of Electronics, University of York, U.K.
(e-mail: rodrigo.delamare@york.ac.uk).}
\thanks{H. V. Poor is with the Department of Electrical Engineering, Princeton University,
Princeton NJ 08544 USA (e-mail: poor@princeton.edu).}
\thanks{Part of this work has been presented at the 2012 Conference on Sensor Signal Processing for Defence, London, UK}}


\maketitle

\begin{abstract}
This paper presents distributed conjugate gradient algorithms for
distributed parameter estimation and spectrum estimation over
wireless sensor networks. In particular, distributed conventional
conjugate gradient (CCG) and modified conjugate gradient (MCG) are
considered, together with incremental and diffusion adaptive
solutions. The distributed CCG and MCG algorithms have an improved
performance in terms of mean square error as compared with
least--mean square (LMS)--based algorithms and a performance that is
close to recursive least--squares (RLS) algorithms. In comparison
with existing centralized or distributed estimation strategies, key
features of the proposed algorithms are: 1) more accurate estimates
and faster convergence speed can be obtained; 2) the design of
preconditioners for CG algorithms, which have the ability to improve
the performance of the proposed CG algorithms is presented and 3)
the proposed algorithms are implemented in the area of distributed
parameter estimation and spectrum estimation. The performance of the
proposed algorithms for distributed estimation is illustrated via
simulations and the resulting algorithms are distributed,
cooperative and able to respond in real time to change in the
environment.
\end{abstract}


\IEEEpeerreviewmaketitle

\section{Introduction}
In recent years, distributed processing has become popular in
wireless communication networks. This approach to processing information consists in collecting data at each node of a network of sensing devices spread over a geographical area, conveying information to the whole network and performing statistical inference in a distributed way \cite{Lopes1}. In this context, for each specific node, a set of neighbor nodes collect their local information and transmit their estimates to a specific node. Then, each specific node combines the collected information together with its local estimate to generate an improved estimate. There are three main protocols for cooperation and exchange of information for distributed processing, incremental, diffusion and consensus strategies, and recent studies indicate that the diffusion strategy is the most effective one \cite{Sayed1}.

In the last recent years, several algorithms have been developed and
reported in the literature for distributed networks.
Steepest-descent, least-mean square (LMS) \cite{Lopes1}, recursive
least squares (RLS) \cite{Sayed13} and affine projection (AP)
\cite{Li} solutions have been considered with incremental adaptive
strategies over distributed networks \cite{Lopes1}, while the LMS,
AP and recursive least squares (RLS) algorithms have been reported
using diffusion adaptive strategies
\cite{Lopes2,Cattivelli2,ls_icassp,ls_camsap,Mateos,ls_journal,dce,abadi,djio_eusipco}.
Although the LMS--based algorithms have their own advantages, when
compared with conjugate gradient (CG) algorithms
\cite{Chang,Chang1,Wang,rrstap}, there are several disadvantages.
First, for the LMS--based algorithms, the adaptation speed is often
slow, especially for the conventional LMS algorithm. Second, with
the increase of the adaptation speed, the system stability may
decrease significantly\cite{Rcdl1}. Furthermore, the RLS--based
algorithms usually have a high computational complexity and are
prone to numerical instability when implemented in hardware
\cite{Haykin}. In order to develop distributed solutions with a more
attractive tradeoff between performance and complexity, we focus on
the development of distributed CG algorithms. To the best of our
knowledge, CG--based algorithms have not been developed so far for
distributed processing. The existing standard CG algorithm has a
faster convergence rate than the LMS-type algorithms and a lower
computational complexity than RLS-type techniques \cite{Axelsson}
even though its performance is often comparable to RLS algorithms.
We consider variants of CG algorithms, including conventional CG
(CCG) algorithm and modified CG (MCG) algorithm.

In this paper, we propose distributed CG algorithms for both
incremental and diffusion adaptive strategies. In particular, we
develop distributed versions of the CCG algorithm and of the MCG
algorithm for use in distributed estimation over sensor networks and
spectrum estimation. The design of preconditioners for CG
algorithms, which have the ability to improve the performance of the
proposed CG algorithms is also presented in this paper. These
algorithms can be widely used in civilian and defence applications,
such as parameter estimation in wireless sensor networks, biomedical
engineering, cellular networks, battlefield information
identification, movement estimation and detection and distributed
spectrum estimation.

In summary, the main contributions of this paper are:
\begin{itemize}
\item We present distributed CG--based algorithms for distributed
estimation that are able to achieve significantly better
performance than existing algorithms.
\item We devise distributed CCG and MCG algorithms with incremental and diffusion adaptive solutions to perform distributed estimation.
\item We implement the proposed CG--based algorithms in the fields of distributed parameter estimation and spectrum estimation.
\item The design of preconditioners for CG algorithms, which have the ability to improve the performance of the proposed CG algorithms is presented.
\item A simulation study of the proposed and existing
distributed estimation algorithms is conducted along with
applications in wireless sensor networks.
\end{itemize}

This paper is organized as follows. Section II describes the system
models. In Section III, the proposed incremental distributed CG--Based algorithms are introduced. We present proposed diffusion distributed CG--Based algorithms in Section IV. The preconditioner design is illustrated in Section V. The
numerical simulation results are provided in Section VI. Finally, we
conclude the paper in Section VII.

Notation: We use boldface upper case letters to denote matrices and
boldface lower case letters to denote vectors. We use $(\cdot)^T$
and $(\cdot)^{-1}$ denote the transpose and inverse operators
respectively, $(\cdot)^H$ for conjugate transposition and
$(\cdot)^*$ for complex conjugate.

\section{System Models}
In this section, we describe the system models of two applications of distributed signal processing, namely, parameter estimation and spectrum estimation.  In these applications, we consider a wireless sensor network which employs distributed signal processing techniques to perform the desired tasks. We consider a set of $N$ nodes, which have limited processing capabilities, distributed over a given geographical area. The nodes are connected and form a network, which is assumed to be partially connected because nodes can exchange information only with neighbors determined by the connectivity topology. We call a network with this property a partially connected network whereas a fully connected network means that data broadcast by a node can be captured by all other nodes in the network in one hop \cite{Bertrand}.

\subsection{Distributed Parameter Estimation}
For distributed parameter estimation, we focus on the processing of an adaptive algorithm for estimating an unknown vector ${\boldsymbol \omega}_o$ with size $M\times1$. The desired signal of each node at time instant $i$ is
\begin{equation}
{d_k{(i)}} = {\boldsymbol {\omega}}_0^H{\boldsymbol x_k{(i)}} +{n_k{(i)}},~~~
i=1,2, \ldots, I ,
\end{equation}
where ${\boldsymbol x_k{(i)}}$ is the $M \times 1$ input signal vector, ${ n_k{(i)}}$ is the Gaussian noise at each node with zero mean and variance $\sigma_{n,k}^2$. At the same time, the output of
the adaptive algorithm for each node is given by
\begin{equation}
{y_k{(i)}} = {\boldsymbol {\omega}_k^H{(i)}}{\boldsymbol x_k{(i)}},~~~ i=1,2,
\ldots, I,
\end{equation}
where ${\boldsymbol \omega_k{(i)}}$ is the local estimate of ${\boldsymbol \omega_0}$
for each node at time instant $ i $.

To compute the optimum solution of the unknown vector, we need to
solve a problem expressed in the form of a minimization of the cost
function in the distributed form for each node $k$:
\begin{equation}\label{Eqn3:cost_function}
{J_{\boldsymbol\omega_k(i)}\big({\boldsymbol \omega_k(i)}\big)} =
{\mathbb{E} \big|{ d_k(i)}- {\boldsymbol \omega_k^H(i)}{\boldsymbol
x_k(i)}\big|^2}
\end{equation}
and the global network cost function could be described as
\begin{equation}
{J_{\boldsymbol\omega}\big({\boldsymbol \omega}\big)} =\sum_{k=1}^N{\mathbb{E} \big|{ d_k(i)}- {\boldsymbol \omega^H}{\boldsymbol
x_k(i)}\big|^2}.
\end{equation}

The optimum solution for the cost function (\ref{Eqn3:cost_function}) is the Wiener solution which is given by
\begin{equation}
\boldsymbol\omega_k(i) =\boldsymbol R_{k}^{-1}(i)\boldsymbol b_{k}(i).
\end{equation}
where the $M\times M$ autocorrelation matrix is given by
${\boldsymbol R_k(i)}=\mathbb{E}[{\boldsymbol x_k(i)} {\boldsymbol x_k^H(i)}]$ and
${\boldsymbol b_k(i)}=\mathbb{E}[{\boldsymbol x_k(i)} {d_k^*(i)}]$ is an $M
\times 1$ cross--correlation matrix. In this paper, we focus on
incremental and diffusion CG--based algorithms to solve the equation
and perform parameter estimation and spectrum estimation in a distributed fashion.

\subsection{Distributed Spectrum Estimation}
In distrusted spectrum estimation, we aim to estimate the spectrum of a transmitted signal $\boldsymbol s$ with N nodes using a wireless sensor network. Let $\boldsymbol\Phi_s(f)$ denote the power spectral density (PSD) of the signal $\boldsymbol s$. The PSD can be represented as a linear combination of some $\mathcal{B}$ basis functions, as described by
\begin{equation}\label{Eqn3:se1}
\Phi_s(f)=\sum\limits_{m=1}^{\mathcal{B}}b_m(f)\omega_{0m}=\boldsymbol b_0^T(f)\boldsymbol\omega_0,
\end{equation}
where $\boldsymbol b_0(f)=[b_1(f),...,b_\mathcal{B}(f)]^T$ is the vector of basis functions evaluated at frequency $f$, $\boldsymbol\omega_0=[\omega_{01},...,\omega_{0\mathcal{B}}]$ is a vector of weighting coefficients representing the power that transmits the signal $\boldsymbol s$ over each basis, and $\mathcal{B}$ is the number of basis functions. For $\mathcal{B}$ sufficiently large, the basis expansion in (\ref{Eqn3:se1}) can well approximate the transmitted spectrum. Possible choices for the set of basis $\{b_m(f)\}_{m=1}^{\mathcal{B}}$ include \cite{Bazerque,Chen22,Zakharov}: rectangular functions, raised cosines, Gaussian bells and Splines.

Let $H_k(f,i)$ be the channel transfer function between a transmit node conveying the signal $\boldsymbol s$ and receive node $k$ at time instant $i$, the PSD of the received signal observed by node k can be expressed as
\begin{align}\label{Eqn3:se2}
I_k(f,i)&=|H_k(f,i)|^2\Phi_s(f)+v^2_{n,k}\notag\\
&=\sum\limits_{m=1}^{\mathcal{B}}|H_k(f,i)|^2b_m(f)\omega_{0m}+v^2_{n,k}\notag\\
&=\boldsymbol b_{k,i}^T(f)\boldsymbol\omega_0+v^2_{n,k}
\end{align}
where $\boldsymbol b_{k,i}^T(f)=[|H_k(f,i)|^2b_m(f)]_{m=1}^{\mathcal{B}}$ and $v^2_{n,k}$ is the receiver noise power at node $k$. For simplification, let us assume that the link between receive node $k$ and the transmit node is perfect and there is no receiver noise at node $k$.

At every time instant $i$, every node $k$ observes measurements of the noisy version of the true PSD $I_k(f,i)$ described by (\ref{Eqn3:se2}) over $N_c$ frequency samples $f_j=f_{min}:(f_{max}-f_{min})/N_c:f_{max}$, for $j=1,...,N_c$, according to the model:
\begin{equation}
d_k^j(i)=\boldsymbol b_{k,i}^T(f_j)\boldsymbol\omega_0+v^2_{n,k}+n_k^j(i).
\end{equation}
The term $n_k^j(i)$ denotes observation noise and have zero mean and variance $\sigma_{n,j}^2$. Collecting measurements over $N_c$ contiguous channels, we obtain a linear model given by
\begin{equation}
\boldsymbol d_k(i)=\boldsymbol B_k(i)\boldsymbol\omega_0+\boldsymbol n_k(i),
\end{equation}
where $\boldsymbol B_k(i)=[\boldsymbol b_{k,i}^T(f_j)]_{j=1}^{N_c}\in \mathbb{R}^{N_c\times\mathcal{B}}$, with $N_c>\mathcal{B}$, and $\boldsymbol n_k(i)$ is a zero mean random vector with covariance matrix $\boldsymbol R_{n,i}$. At this point, we can generate the cost function for node $k$ as:
\begin{equation}\label{Eqn3:cf_se}
{J_{\boldsymbol \omega_k(i)}({\boldsymbol \omega_k(i)})} = {\mathbb{E} \big|{\boldsymbol d_k(i)}-\boldsymbol B_k(i)\boldsymbol\omega_k(i)\big|^2}
\end{equation}
and the global network cost function could be described as
\begin{equation}\label{Eqn3:cf_se2}
{J_{\boldsymbol\omega}\big({\boldsymbol \omega}\big)} =\sum_{k=1}^N{\mathbb{E} \big|{\boldsymbol d_k(i)}- \boldsymbol B_k(i)\boldsymbol \omega\big|^2}.
\end{equation}

\section{Proposed Incremental Distributed CG--Based Algorithms}

In this section, we propose two CG--based algorithms which are based
on the CCG \cite{Chang} and MCG \cite{Chang1} algorithms with
incremental distributed solution for distributed parameter
estimation and spectrum estimation over wireless sensor networks.
Other reduced-rank techniques
\cite{int,jio,mwfccm,jiols,jiomimo,Rcdl1,wlmwf} and distributed
strategies \cite{armo} can also be considered. Before we present the
proposed incremental distributed CG--Based algorithms, we introduce
the basic CG algorithm in detail.

\subsection{The Conjugate Gradient (CG) Algorithm}

The CG algorithm is well known for its faster convergence rate than
the LMS algorithm and lower computational complexity than the RLS
algorithm \cite{Axelsson,Golub,Chang}. In adaptive filtering
techniques, the CG algorithm applied to the system $\boldsymbol
R\boldsymbol\omega=\boldsymbol b$, starts with an initial guess of
the solution $\boldsymbol\omega(0)$, with an initial residual
$\boldsymbol g(0)=\boldsymbol b$, and with an initial search
direction that is equal to the initial residual: $\boldsymbol
p(0)=\boldsymbol g(0)$, where $\boldsymbol R$ is the correlation or
the covariance matrix of the input signal and $\boldsymbol b$ is the
cross--correlation vector between the desired signal and the input
signal.

The strategy for the conjugate gradient method is that at step $j$, the residual $\boldsymbol g(j)=\boldsymbol b-\boldsymbol R\boldsymbol\omega(j)$ is orthogonal to the Krylov subspace generated by $\boldsymbol b$, and therefore each residual is perpendicular to all the previous residuals. The residual is computed at each step.

The solution at the next step is achieved using a search direction that is only a linear combination of the previous search directions, which for $\boldsymbol\omega(1)$ is just a combination between the previous and the current residual.

Then, the solution at step $j$, $\boldsymbol\omega(j)$, could be obtained through $\boldsymbol\omega(j-1)$ from the previous iteration plus a step size $\alpha(j)$ times the last search direction. The immediate benefit of the search directions is that there is no need to store the previous search directions. Using the orthogonality of the residuals to these previous search directions, the search is linearly independent of the previous directions. For the solution in the next step, a new search direction is computed, as well as a new residual and new step size. To provide an optimal approximate solution of $\boldsymbol\omega$, the step size $\alpha(j)$ is calculated according to \cite{Axelsson,Golub,Chang}.

\begin{figure}[!htb]
\begin{center}
\def\epsfsize#1#2{0.5\columnwidth}
\epsfbox{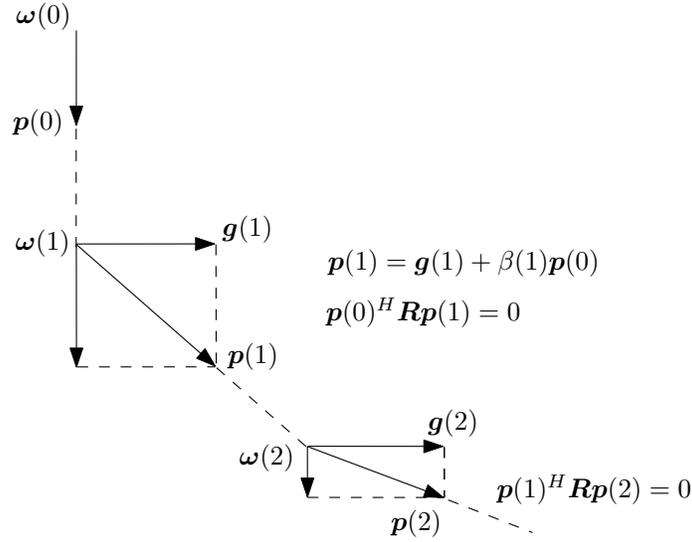}\caption{\footnotesize Searching Direction of the
CG Algorithm} \label{fig2:CG_algorithm}
\end{center}
\end{figure}

To illustrate the CG algorithm, Fig. \ref{fig2:CG_algorithm} shows
how the CG algorithm finds the approximate solution to the exact
solution. The iterative formulas of the CG algorithm
\cite{Axelsson,Golub,Chang} are concluded in Table
\ref{table2:CG_algorithm}.

\begin{table}[!htb]
\centering \caption{Main Steps for CG algorithm.}
\begin{tabular}{l}
\hline
--\ Step size:$\alpha(j)=\frac{\boldsymbol g(j-1)^H\boldsymbol g(j-1)}{\boldsymbol p(j-1)^H\boldsymbol R\boldsymbol p(j-1)}$\\
--\ Approximate solution: $\boldsymbol\omega(j)=\boldsymbol\omega(j-1)+\alpha(j)\boldsymbol p(j-1)$\\
--\ Residual: $\boldsymbol g(j)=\boldsymbol g(j-1)-\alpha(j)\boldsymbol R\boldsymbol p(j-1)$\\
--\ Improvement at step $i$: $\beta(j)=\frac{\boldsymbol g(j)^H\boldsymbol g(j)}{\boldsymbol g(j-1)^H\boldsymbol g(j-1)}$\\
--\ Search direction: $\boldsymbol p(j)=\boldsymbol g(j)+\beta(j)\boldsymbol p(j-1)$\\
\hline
\end{tabular}
\label{table2:CG_algorithm}
\end{table}

\subsection{Incremental Distributed CG--Based Solutions}
\begin{figure}[!htb]
\begin{center}
\def\epsfsize#1#2{0.7\columnwidth}
\epsfbox{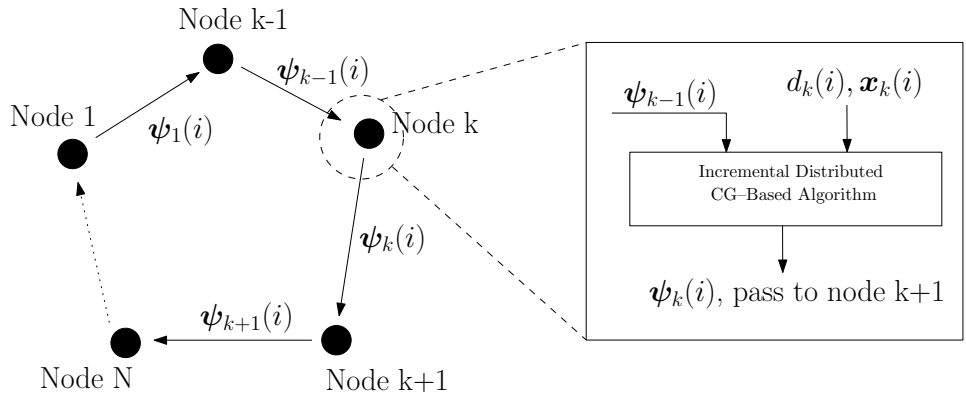} \caption{\footnotesize Incremental distributed
CG--based network processing} \label{fig3:IDCG}
\end{center}
\end{figure}

In the incremental distributed strategy, each node is only allowed to communicate with its direct neighbor at each time
instant. To describe the whole process, we define a cycle where each
node in this network could only access its immediate neighbor in
this cycle \cite{Lopes1}. The quantity $\boldsymbol\psi_k {(i)}$ is
defined as a local estimate of the unknown vector $\boldsymbol\omega_0$ at time instant $i$. As
a result, we assume that node $k$ has access to an estimate of
$\boldsymbol\omega_0$  at its immediate neighbor node $k-1$ which is
$\boldsymbol\psi_{k-1} {(i)}$ in the defined cycle. Fig.\ref{fig3:IDCG}
illustrates this processing.

In the following, we introduce two kinds of incremental distributed
CG--based algorithms, which are the incremental distributed CCG
(IDCCG) algorithm and  the incremental distributed MCG (IDMCG)
algorithm.

\subsubsection{Proposed IDCCG Algorithm}

Based on the main steps of CG algorithm which are described in
Table. \ref{table2:CG_algorithm}, we introduce the main steps of the
proposed IDCCG algorithm. In the IDCCG algorithm, the iteration
procedure is introduced. At the $j$th iteration of time instant $i$,
the step size $\alpha_k^j{(i)}$ for updating the local estimate at
node $k$ is defined as:
\begin{equation}\label{Eqn3:alpha}
{\alpha_k^j{(i)}} = \frac{\big(\boldsymbol g_k^{j-1}{(i)}\big)^H {\boldsymbol g_k^{j-1}{(i)}}}
{\big(\boldsymbol p_k^{j-1}{(i)}\big)^H{\boldsymbol R_k{(i)}} {\boldsymbol p_k^{j-1}{(i)}}} ,
\end{equation}
where $\boldsymbol p_k(i)$ is the search direction and defined as
\begin{equation}\label{Eqn3:cg1}
{\boldsymbol p_k^j{(i)}} = {\boldsymbol g_k^j{(i)}} + {\beta_k^j{(i)}}{\boldsymbol p_k^{j-1}{(i)}}.
\end{equation}
In (\ref{Eqn3:cg1}), the coefficient $\beta_k^j{(i)}$ is calculated by the Gram--Schmidt orthogonalization procedure \cite{Golub} for the conjugacy:
\begin{equation}
{\beta_k^j{(i)}} = \frac{ \big(\boldsymbol g_k^j{(i)}\big)^H {\boldsymbol g_k^j{(i)}}}{ \big(\boldsymbol g_k^{j-1}{(i)}\big)^H {\boldsymbol g_k^{j-1}{(i)}}}. \end{equation}
$\boldsymbol g_k^j{(i)}$ is the residual, which is obtained as
\begin{equation}
{\boldsymbol g_k^j{(i)}}=\boldsymbol g_k^{j-1}{(i)} - \alpha_k^j{(i)}{\boldsymbol R_k{(i)}}\boldsymbol p_k^{j-1}{(i)}.
\end{equation}
The initial search direction is equal to the initial residual, which is given by $\boldsymbol p_k^0{(i)}=\boldsymbol g_k^0{(i)}=\boldsymbol b_k{(i)}-\boldsymbol R_k{(i)}\boldsymbol \psi_k^0(i)$. Then, the local estimate is updated as
\begin{equation}
{\boldsymbol \psi_k^j{(i)}} = {\boldsymbol \psi_{k}^{j-1}{(i)}} + {\alpha_k^j{(i)}}{\boldsymbol p_k^{j-1}{(i)}}.
\end{equation}

There are two ways to compute the
correlation and cross--correlation matrices which are the 'finite
sliding data window' and the 'exponentially decaying data
window' \cite{Chang}. In this paper, we mainly focus on the
'exponentially decaying data window'. The recursions are
given by:
\begin{equation}\label{Eqn3:R}
{\boldsymbol R_k{(i)}} = {\lambda_f}{\boldsymbol R_{k}{(i-1)}}+ {\boldsymbol
x_k{(i)}}{\boldsymbol x_k^H{(i)}}
\end{equation}
and
\begin{equation}\label{Eqn3:b}
{\boldsymbol b_k{(i)}} = {\lambda_f}{\boldsymbol b_{k}{(i-1)}} +
d_k^*{(i)}{\boldsymbol x_k{(i)}}
\end{equation}
where $\lambda_f$ is the forgetting factor. The IDCCD algorithm is summarized in Table \ref{table3:IDCCG}
\begin{table}
\caption{IDCCG Algorithm} \centering
\begin{tabular}{l}
\hline
Initialization:\\
$\boldsymbol \omega(0)=\boldsymbol 0$\\
For each time instant $i$=1,2, \ldots, I\\
\ \ \ \ \ \ \ \ $\boldsymbol \psi_1^0(i)=\boldsymbol \omega(i-1)$\\
\ \ \ \ \ \ \ \ For each node $k$=1,2, \ldots, N\\
\ \ \ \ \ \ \ \ \ \ \ \ \ \ \ \ ${\boldsymbol R_k{(i)}} = {\lambda_f}{\boldsymbol R_{k}{(i-1)}} + {\boldsymbol x_k{(i)}}{ {\boldsymbol x_k^H{(i)}}} \notag$\\
\ \ \ \ \ \ \ \ \ \ \ \ \ \ \ \ ${\boldsymbol b_k{(i)}} = {\lambda_f}{\boldsymbol b_{k}{(i-1)}}+ { d_k^*{(i)}}{\boldsymbol x_k{(i)}} \notag$\\
\ \ \ \ \ \ \ \ \ \ \ \ \ \ \ \ $\boldsymbol p_k^0{(i)}=\boldsymbol g_k^0{(i)}=\boldsymbol b_k{(i)}-\boldsymbol R_k{(i)}\boldsymbol \psi_k^0(i)$\\
\ \ \ \ \ \ \ \ \ \ \ \ \ \ \ \ For iterations $j$=1,2, \ldots, J\\
\ \ \ \ \ \ \ \ \ \ \ \ \ \ \ \ \ \ \ \ \ \ \ \ ${\alpha_k^j{(i)}} = \frac{\big(\boldsymbol g_k^{j-1}{(i)}\big)^H {\boldsymbol g_k^{j-1}{(i)}}}
{\big(\boldsymbol p_k^{j-1}{(i)}\big)^H{\boldsymbol R_k{(i)}} {\boldsymbol p_k^{j-1}{(i)}}}$\\
\ \ \ \ \ \ \ \ \ \ \ \ \ \ \ \ \ \ \ \ \ \ \ \ ${\boldsymbol \psi_k^j{(i)}} = {\boldsymbol \psi_{k}^{j-1}{(i)}} + {\alpha_k^j{(i)}}{\boldsymbol p_k^{j-1}{(i)}}$\\
\ \ \ \ \ \ \ \ \ \ \ \ \ \ \ \ \ \ \ \ \ \ \ \ ${\boldsymbol g_k^j{(i)}}=\boldsymbol g_k^{j-1}{(i)} - \alpha_k^j{(i)}{\boldsymbol R_k{(i)}}\boldsymbol p_k^{j-1}{(i)}$\\
\ \ \ \ \ \ \ \ \ \ \ \ \ \ \ \ \ \ \ \ \ \ \ \ ${\beta_k^j{(i)}} = \frac{ \big(\boldsymbol g_k^j{(i)}\big)^H {\boldsymbol g_k^j{(i)}}}{ \big(\boldsymbol g_k^{j-1}{(i)}\big)^H {\boldsymbol g_k^{j-1}{(i)}}}$\\
\ \ \ \ \ \ \ \ \ \ \ \ \ \ \ \ \ \ \ \ \ \ \ \ ${\boldsymbol p_k^j{(i)}} = {\boldsymbol g_k^j{(i)}} + {\beta_k^j{(i)}}{\boldsymbol p_k^{j-1}{(i)}}$\\
\ \ \ \ \ \ \ \ \ \ \ \ \ \ \ \ End\\
\ \ \ \ \ \ \ \ \ \ \ \ \ \ \ \ When $k<N$\\
\ \ \ \ \ \ \ \ \ \ \ \ \ \ \ \ $\boldsymbol\psi_{k+1}^0{(i)} = \boldsymbol\psi_k^J(i)$\\
\ \ \ \ \ \ \ \ End\\
\ \ \ \ \ \ \ \ $\boldsymbol \omega(i)=\boldsymbol\psi_N^J{(i)}$\\
End\\
\hline
\end{tabular}
\label{table3:IDCCG}
\end{table}

\subsubsection{Proposed IDMCG Algorithm}

The idea of the IDMCG algorithm comes from the existing CCG
algorithm. For the IDMCG solution, a recursive formulation for the
residual vector is employed, which can be found by using
(\ref{Eqn3:alpha}), (\ref{Eqn3:R}) and (\ref{Eqn3:b})
\cite{Chang,Wang}, resulting in
\begin{equation}\label{Eqn3:g_k}
\begin{split}
{\boldsymbol g_k{(i)}} &= {\boldsymbol b_k{(i)}} - {\boldsymbol R_k{(i)}}{\boldsymbol \psi_k{(i)}} \\
& = {\lambda_f}{\boldsymbol g_{k}{(i-1)}} - {\alpha_k{(i)}}{\boldsymbol R_k{(i)}}{\boldsymbol p_k{(i-1)}}
+ {\boldsymbol x_k{(i)}}[{d_k{(i)}}-{\boldsymbol \psi_{k-1}^H{(i)}}{\boldsymbol x_k{(i)}}].
\end{split}
\end{equation}
Premultiplying (\ref{Eqn3:g_k}) by $\boldsymbol p_k^H(i-1)$ gives
\begin{equation}
\begin{split}
\boldsymbol p_k^H(i-1){\boldsymbol g_k{(i)}}&= {\lambda_f}\boldsymbol p_k^H(i-1){\boldsymbol g_{k}{(i-1)}} - {\alpha_k{(i)}}\boldsymbol p_k^H(i-1){\boldsymbol R_k{(i)}}{\boldsymbol p_k{(i-1)}}\\
& \ \ \ \ +\boldsymbol p_k^H(i-1){\boldsymbol x_k{(i)}}[{d_k{(i)}}-{\boldsymbol \psi_{k-1}^H{(i)}}{\boldsymbol x_k{(i)}}].
\end{split}
\end{equation}
Taking the expectation of both sides and considering $\boldsymbol p_k(i-1)$ uncorrelated with $\boldsymbol x_k(i)$, $ d_k(i)$ and
$\boldsymbol \psi_{k-1}(i)$ yields
\begin{equation}\label{Eqn3:e_p_k}
\begin{split}
\mathbb{E}[{\boldsymbol p_k^H{(i-1)}}{\boldsymbol g_k{(i)}}] & \approx {\lambda_f}\mathbb{E}[{\boldsymbol p_k{(i-1)}}^H{\boldsymbol g_{k-1}{(i)}}]
 - \mathbb{E}[{\alpha_k{(i)}}]\mathbb{E}[{\boldsymbol p_k^H{(i-1)}}{\boldsymbol R_k{(i)}}{\boldsymbol p_k}{(i-1)}]
\\
& \ \ \ \ + \mathbb{E}[{\boldsymbol p_k^H{(i-1)}}]\mathbb{E}\big[{\boldsymbol x_k{(i)}}[{d_k{(i)}}-{\boldsymbol \omega_{k-1}^H{(i)}}{\boldsymbol x_k{(i)}}]\big].
\end{split}
\end{equation}
Assuming that the algorithm converges, the last term of (\ref{Eqn3:e_p_k}) could be neglected and we will obtain:
\begin{equation}
\mathbb{E}[{\alpha_k}{(i)}] = \frac{\mathbb{E}[{\boldsymbol p_{k}^H{(i-1)}}{\boldsymbol g_{k}}{(i)}]-{\lambda_f}\mathbb{E}[{\boldsymbol p_{k}^H{(i-1)}}{\boldsymbol g_{k}{(i-1)}}]}{\mathbb{E}[{\boldsymbol p_k^H{(i-1)}}{\boldsymbol R_k{(i)}}{\boldsymbol p_k{(i-1)}}]}
\end{equation}
and
\begin{equation}\label{Eqn3:inequalities}
\begin{split}
({\lambda_f-0.5})\frac{\mathbb{E}[{\boldsymbol p_{k}^H{(i-1)}}{\boldsymbol g_{k}{(i-1)}}]}
{\mathbb{E}[{\boldsymbol p_k^H{(i-1)}}{\boldsymbol R_k{(i)}}{\boldsymbol p_k{(i-1)}}]}
\leq \mathbb{E}[{\alpha_k{(i)}}]
\leq \frac{\mathbb{E}[{\boldsymbol p_{k}^H{(i-1)}}{\boldsymbol g_{k}{(i-1)}}]}
{\mathbb{E}[{\boldsymbol p_k^H{(i-1)}}{\boldsymbol R_k{(i)}}{\boldsymbol p_k{(i-1)}}]}
\end{split}
\end{equation}
The inequalities in (\ref{Eqn3:inequalities}) are satisfied if we define \cite{Chang}:
\begin{equation}
{\alpha_k{(i)}} = {\eta}\frac{{\boldsymbol p_{k}^H{(i-1)}}{\boldsymbol g_{k}{(i-1)}}}
{{\boldsymbol p_k^H{(i-1)}}{\boldsymbol R_k{(i)}}{\boldsymbol p_k{(i-1)}}},
\end{equation}
where $(\lambda_f -0.5)\leq\eta\leq\lambda_f $.
The direction vector $\boldsymbol p_k(i)$ for the IDMCG algorithm is defined by
\begin{equation}
{\boldsymbol p_{k}{(i)}} = {\boldsymbol g_k{(i)}} + {\beta_k{(i)}}{\boldsymbol p_k{(i-1)}}.
\end{equation}
For the IDMCG algorithm, for the computation of $\beta_k{(i)}$, the Polak--Ribiere method \cite{Chang}, which is given by
\begin{equation}
{\beta_k{(i)}} = \frac{\big({\boldsymbol g_k{(i)}-\boldsymbol g_{k}^H{(i-1)}\big)\boldsymbol g_k{(i)}}}
{{{\boldsymbol g_{k}^H{(i-1)}}}{\boldsymbol g_{k}{(i-1)}}}
\end{equation}
should be used for improved performance, according to \cite{Fletcher,Shanno}.

In the comparison of the IDCCG algorithm with the IDMCG algorithm, the difference between these two strategies is that IDCCG needs to run $J$ iterations while IDMCG only needs one iteration. The details of the IDMCG solution is shown in Table \ref{table3:IDMCG}.
\begin{table}
\caption{IDMCG Algorithm} \centering
\begin{tabular}{l}
\hline
Initialization:\\
$\boldsymbol \omega(0)=\boldsymbol 0$\\
For each node $k$=1,2, \ldots, N\\
\ \ \ \ \ \ \ \ ${\boldsymbol b_k{(1)}} = { d_k^*{(1)}}{\boldsymbol x_k{(1)}} $\\
\ \ \ \ \ \ \ \ $\boldsymbol p_k{(0)}=\boldsymbol g_k{(0)}=\boldsymbol b_k{(1)}$\\
End\\
For each time instant $i$=1,2, \ldots, I\\
\ \ \ \ \ \ \ \ $\boldsymbol \psi_0(i)=\boldsymbol \omega(i-1)$\\
\ \ \ \ \ \ \ \ For each node $k$=1,2, \ldots, N\\
\ \ \ \ \ \ \ \ \ \ \ \ \ \ \ \ ${\boldsymbol R_k{(i)}} = {\lambda_f}{\boldsymbol R_{k}{(i-1)}} + {\boldsymbol x_k{(i)}}{ {\boldsymbol x_k^H{(i)}}} \notag$\\
\ \ \ \ \ \ \ \ \ \ \ \ \ \ \ \ ${\alpha_k{(i)}} = {\eta}\frac{{\boldsymbol p_{k}^H{(i-1)}}{\boldsymbol g_{k}{(i-1)}}}
{{\boldsymbol p_k^H{(i-1)}}{\boldsymbol R_k{(i)}}{\boldsymbol p_k{(i-1)}}}$\\
\ \ \ \ \ \ \ \ \ \ \ \ \ \ \ \ where $(\lambda_f -0.5)\leq\eta\leq\lambda_f $\\
\ \ \ \ \ \ \ \ \ \ \ \ \ \ \ \ ${\boldsymbol \psi_k{(i)}} = {\boldsymbol \psi_{k-1}{(i)}} + {\alpha_k{(i)}}{\boldsymbol p_k{(i-1)}}$\\
\ \ \ \ \ \ \ \ \ \ \ \ \ \ \ \ ${\boldsymbol g_k{(i)}} = {\lambda_f}{\boldsymbol g_{k}{(i-1)}} - {\alpha_k{(i)}}{\boldsymbol R_k{(i)}}{\boldsymbol p_k{(i-1)}}+ {\boldsymbol x_k{(i)}}[{d_k{(i)}}-{\boldsymbol \psi_{k-1}^H{(i)}}{\boldsymbol x_k{(i)}}]$\\
\ \ \ \ \ \ \ \ \ \ \ \ \ \ \ \ ${\beta_k{(i)}} = \frac{\big({\boldsymbol g_k{(i)}-\boldsymbol g_{k}^H{(i-1)}\big)\boldsymbol g_k{(i)}}}
{{{\boldsymbol g_{k}^H{(i-1)}}}{\boldsymbol g_{k}{(i-1)}}}$\\
\ \ \ \ \ \ \ \ \ \ \ \ \ \ \ \ ${\boldsymbol p_k{(i)}} = {\boldsymbol g_k{(i)}} + {\beta_k{(i)}}{\boldsymbol p_k{(i-1)}}$\\
\ \ \ \ \ \ \ \ End\\
\ \ \ \ \ \ \ \ $\boldsymbol \omega(i)=\boldsymbol\psi_N{(i)}$\\
End\\
\hline
\end{tabular}
\label{table3:IDMCG}
\end{table}

\subsection{Computational Complexity}

To analyze the proposed incremental distributed CG algorithms, we
detail the computational complexity in terms of arithmetic
operations. Additions and multiplications are used to measure the
complexity and are listed in Table \ref{table3:cc1}. The parameter
$M$ is the length of the unknown vector $\boldsymbol\omega_0$ that
needs to be estimated. It is obvious that the complexity of the
IDCCG solution depends on the number of iterations $J$ and an
advantage of the IDMCG algorithm is that it only requires one
iteration per time instant.
\begin{table}
\centering \caption{Computational Complexity of Different Incremental Algorithms}
\begin{tabular}{|c|c|c|}
\hline
Algorithm&Additions&Multiplications\\
\hline
IDCCG&$M^2+M$&$2M^2+2M$\\
&$+J(M^2+6M-4)$&$J(M^2+7M+3)$\\
\hline
IDMCG&$2M^2+10M-4$&$3M^2+12M+3$\\
\hline
Incremental LMS \cite{Lopes1} &$4M-1$&$3M+1$\\
\hline
Incremental RLS \cite{Lopes1}&$4M^2+12M+1$&$4M^2+12M-1$\\
\hline
\end{tabular}
\label{table3:cc1}
\end{table}

\section{Proposed Diffusion Distributed CG--Based Algorithms}

In this section, we detail the proposed diffusion distributed CCG
(DDCCG) and diffusion distributed MCG (DDMCG) algorithms for
distributed parameter estimation and spectrum estimation using
wireless sensor networks.

\subsection{Diffusion Distributed CG--Based Algorithms}

\begin{figure}[!htb]
\begin{center}
\def\epsfsize#1#2{0.7\columnwidth}
\epsfbox{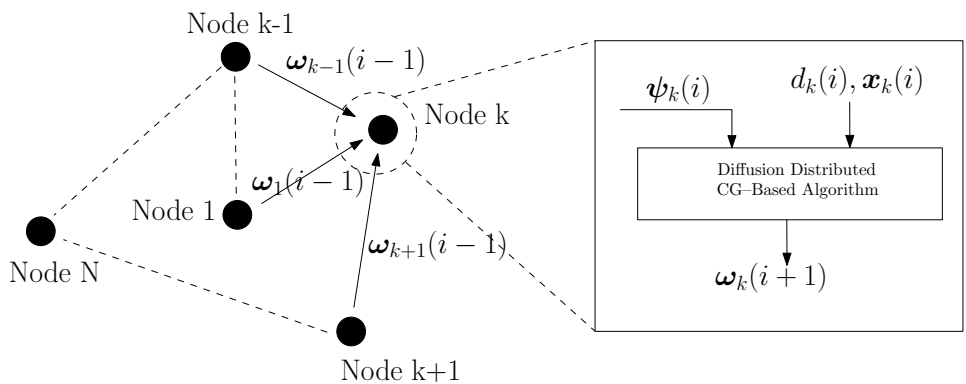} \caption{\footnotesize Diffusion Distributed
CG--Based Network Processing}\vspace{-0.5em} \label{fig3:DDCG}
\end{center}
\end{figure}

In the derivation of diffusion distributed CG--based strategy, we
consider a network structure where each node from the same
neighborhood could exchange information with each other at every
time instant. For each node in the network, the CTA scheme
\cite{Cattivelli3} is employed. Each node can collect information
from all its neighbors and itself, and then convey all the
information to its local adaptive algorithm and update the estimate
of the weight vector through our algorithms. Specifically, at any
time instant $i$, we define that node $k$ has access to a set of
estimates $\{\boldsymbol\omega_l(i-1)\}_{l\in \mathcal{N}_k}$ from
its neighbors, where $N_k$ denotes the set of neighbor nodes of node
$k$ including node $k$ itself. Then, these local estimates are
combined at node $k$ as
\begin{equation}\label{Eqn3:c}
{\boldsymbol\psi_k(i)} = \sum_{l\in \mathcal{N}_k} c_{kl} \boldsymbol\omega_l(i-1)
\end{equation}
where $c_{kl}$ is the combining coefficient. There are many ways to calculate the combining
coefficient $c_{kl}$ which include the Hastings \cite{Zhao1}, the
Metropolis \cite{Xiao}, the Laplacian \cite{Olfati} and the nearest
neighbor \cite{Jadbabaie} rules. In this paper, due to its simplicity
and good performance \cite{Zhao1} we adopt the Metropolis rule given
by
\begin{equation}
c_{kl}=\left\{\begin{array}{ll}
\frac{1}
{max\{|\mathcal{N}_k|,|\mathcal{N}_l|\}},\ \ $if\  $k\neq l$\ are linked$\\
1 - \sum\limits_{l\in \mathcal{N}_k / k} c_{kl}, \ \ $for\  $k$\ =\ $l$$,
\end{array}
\right.
\end{equation}
where $|\mathcal{N}_k|$ denotes the cardinality of $\mathcal{N}_k$. The combining coefficients $c_{kl}$ should satisfy
\begin{equation}
\sum\limits_{l\in \mathcal{N}_k \forall k} c_{kl} =1.
\end{equation}
For the proposed diffusion distributed CG--based algorithms, the whole processing is shown in Fig. \ref{fig3:DDCG}.

\subsubsection{Proposed DDCCG Algorithm}

For the DDCCG algorithm, (\ref{Eqn3:c}) is employed to combine the
estimates $\boldsymbol\omega_l{(i-1)}, l\in \mathcal{N}_k$ from node
$k$'s neighbor nodes and then the estimate at node $k$ is updated
as:
\begin{equation}
{\boldsymbol \omega_k^j{(i)}} = {\boldsymbol \omega_{k}^{j-1}{(i)}} + {\alpha_k^j{(i)}}{\boldsymbol p_k^{j-1}{(i)}},
\end{equation}
where $\boldsymbol \omega_{k}^0(i)=\boldsymbol\psi_k{(i)}$. The rest of the derivation is similar to the IDCCG solution and the pseudo--code is detailed in Table \ref{table3:DDCCG}.
\begin{table}
\centering \caption{DDCCG Algorithm}
\begin{tabular}{l}
\hline
Initialization:\\
$\boldsymbol \omega_k(0)=\boldsymbol 0, k$=1,2, \ldots, N\\
For each time instant $i$=1,2, \ldots, I\\
\ \ \ \ \ \ \ \ For each node $k$=1,2, \ldots, N (Combination Step)\\
\ \ \ \ \ \ \ \ \ \ \ \ \ \ \ \ ${\boldsymbol\psi_k(i)} = \sum_{l\in \mathcal{N}_k} c_{kl} \boldsymbol\omega_l(i-1)$\\
\ \ \ \ \ \ \ \ End\\
\ \ \ \ \ \ \ \ For each node $k$=1,2, \ldots, N (Adaptation Step)\\
\ \ \ \ \ \ \ \ \ \ \ \ \ \ \ \ ${\boldsymbol R_k{(i)}} = {\lambda_f}{\boldsymbol R_{k}{(i-1)}} + {\boldsymbol x_k{(i)}}{ {\boldsymbol x_k^H{(i)}}} \notag$\\
\ \ \ \ \ \ \ \ \ \ \ \ \ \ \ \ ${\boldsymbol b_k{(i)}} = {\lambda_f}{\boldsymbol b_{k}{(i-1)}}+ { d_k^*{(i)}}{\boldsymbol x_k{(i)}} \notag$\\
\ \ \ \ \ \ \ \ \ \ \ \ \ \ \ \ $\boldsymbol \omega_k^0(i)=\boldsymbol \psi_k(i)$\\
\ \ \ \ \ \ \ \ \ \ \ \ \ \ \ \ $\boldsymbol p_k^0{(i)}=\boldsymbol g_k^0{(i)}=\boldsymbol b_k{(i)}-\boldsymbol R_k{(i)}\boldsymbol \omega_k^0(i)$\\
\ \ \ \ \ \ \ \ \ \ \ \ \ \ \ \ For iterations $j$=1,2, \ldots, J\\
\ \ \ \ \ \ \ \ \ \ \ \ \ \ \ \ \ \ \ \ \ \ \ \ ${\alpha_k^j{(i)}} = \frac{\big(\boldsymbol g_k^{j-1}{(i)}\big)^H {\boldsymbol g_k^{j-1}{(i)}}}
{\big(\boldsymbol p_k^{j-1}{(i)}\big)^H{\boldsymbol R_k{(i)}} {\boldsymbol p_k^{j-1}{(i)}}}$\\
\ \ \ \ \ \ \ \ \ \ \ \ \ \ \ \ \ \ \ \ \ \ \ \ ${\boldsymbol \omega_k^j{(i)}} = {\boldsymbol \omega_{k}^{j-1}{(i)}} + {\alpha_k^j{(i)}}{\boldsymbol p_k^{j-1}{(i)}}$\\
\ \ \ \ \ \ \ \ \ \ \ \ \ \ \ \ \ \ \ \ \ \ \ \ ${\boldsymbol g_k^j{(i)}}=\boldsymbol g_k^{j-1}{(i)} - \alpha_k^j{(i)}{\boldsymbol R_k{(i)}}\boldsymbol p_k^{j-1}{(i)}$\\
\ \ \ \ \ \ \ \ \ \ \ \ \ \ \ \ \ \ \ \ \ \ \ \ ${\beta_k^j{(i)}} = \frac{ \big(\boldsymbol g_k^j{(i)}\big)^H {\boldsymbol g_k^j{(i)}}}{ \big(\boldsymbol g_k^{j-1}{(i)}\big)^H {\boldsymbol g_k^{j-1}{(i)}}}$\\
\ \ \ \ \ \ \ \ \ \ \ \ \ \ \ \ \ \ \ \ \ \ \ \ ${\boldsymbol p_k^j{(i)}} = {\boldsymbol g_k^j{(i)}} + {\beta_k^j{(i)}}{\boldsymbol p_k^{j-1}{(i)}}$\\
\ \ \ \ \ \ \ \ \ \ \ \ \ \ \ \ End\\
\ \ \ \ \ \ \ \ \ \ \ \ \ \ \ \ $\boldsymbol\omega_k{(i)} = \boldsymbol\omega_k^J(i)$\\
\ \ \ \ \ \ \ \ End\\
End\\
\hline
\end{tabular}
\label{table3:DDCCG}
\end{table}

\subsubsection{Proposed DDMCG Algorithm}

For the DDMCG algorithm, the iteration $j$ is removed and the estimate at node $k$ is updated as:
\begin{equation}
{\boldsymbol \omega_k{(i)}} = {\boldsymbol \psi_{k}{(i)}} + {\alpha_k{(i)}}{\boldsymbol p_k{(i)}},
\end{equation}
The complete DDMCG solution is described in Table \ref{table3:DDMCG}.
\begin{table}
\centering \caption{DDMCG Algorithm}
\begin{tabular}{l}
\hline
Initialization:\\
$\boldsymbol \omega_k(0)=\boldsymbol 0, k$=1,2, \ldots, N\\
For each node $k$=1,2, \ldots, N\\
\ \ \ \ \ \ \ \ ${\boldsymbol b_k{(1)}} = { d_k^*{(1)}}{\boldsymbol x_k{(1)}} $\\
\ \ \ \ \ \ \ \ $\boldsymbol p_k{(0)}=\boldsymbol g_k{(0)}=\boldsymbol b_k{(1)}$\\
End\\
For each time instant $i$=1,2, \ldots, I\\
\ \ \ \ \ \ \ \ For each node $k$=1,2, \ldots, N (Combination Step)\\
\ \ \ \ \ \ \ \ \ \ \ \ \ \ \ \ ${\boldsymbol\psi_k(i)} = \sum_{l\in \mathcal{N}_k} c_{kl} \boldsymbol\omega_l(i-1)$\\
\ \ \ \ \ \ \ \ End\\
\ \ \ \ \ \ \ \ For each node $k$=1,2, \ldots, N (Adaptation Step)\\
\ \ \ \ \ \ \ \ \ \ \ \ \ \ \ \ ${\boldsymbol R_k{(i)}} = {\lambda_f}{\boldsymbol R_{k}{(i-1)}} + {\boldsymbol x_k{(i)}}{ {\boldsymbol x_k^H{(i)}}} \notag$\\
\ \ \ \ \ \ \ \ \ \ \ \ \ \ \ \ ${\boldsymbol b_k{(i)}} = {\lambda_f}{\boldsymbol b_{k}{(i-1)}}+ { d_k^*{(i)}}{\boldsymbol x_k{(i)}} \notag$\\
\ \ \ \ \ \ \ \ \ \ \ \ \ \ \ \ ${\alpha_k{(i)}} = {\eta}\frac{{\boldsymbol p_{k}^H{(i-1)}}{\boldsymbol g_{k}{(i-1)}}}
{{\boldsymbol p_k^H{(i-1)}}{\boldsymbol R_k{(i)}}{\boldsymbol p_k{(i-1)}}}$\\
\ \ \ \ \ \ \ \ \ \ \ \ \ \ \ \ where $(\lambda_f -0.5)\leq\eta\leq\lambda_f $\\

\ \ \ \ \ \ \ \ \ \ \ \ \ \ \ \ ${\boldsymbol \omega_k{(i)}} = {\boldsymbol \psi_{k}{(i)}} + {\alpha_k{(i)}}{\boldsymbol p_k{(i-1)}}$\\
\ \ \ \ \ \ \ \ \ \ \ \ \ \ \ \ ${\boldsymbol g_k{(i)}} = {\lambda_f}{\boldsymbol g_{k}{(i-1)}} - {\alpha_k{(i)}}{\boldsymbol R_k{(i)}}{\boldsymbol p_k{(i-1)}}+ {\boldsymbol x_k{(i)}}[{d_k{(i)}}-{\boldsymbol \psi_{k-1}^H{(i)}}{\boldsymbol x_k{(i)}}]$\\
\ \ \ \ \ \ \ \ \ \ \ \ \ \ \ \ ${\beta_k{(i)}} = \frac{\big({\boldsymbol g_k{(i)}-\boldsymbol g_{k}^H{(i-1)}\big)\boldsymbol g_k{(i)}}}
{{{\boldsymbol g_{k}^H{(i-1)}}}{\boldsymbol g_{k}{(i-1)}}}$\\
\ \ \ \ \ \ \ \ \ \ \ \ \ \ \ \ ${\boldsymbol p_k{(i)}} = {\boldsymbol g_k{(i)}} + {\beta_k{(i)}}{\boldsymbol p_k{(i-1)}}$\\
\ \ \ \ \ \ \ \ End\\
End\\
\hline
\end{tabular}
\label{table3:DDMCG}
\end{table}

\subsection{Computational Complexity}

The computational complexity is used to analyse the proposed
diffusion distributed CG--based algorithms where additions and
multiplications are measured. The details are listed in Table
\ref{table3:cc2}. Similarly to the incremental distributed CG--based
algorithms, it is clear that the complexity of the DDCCG solution
depends on the iteration number $J$ and both DDCCG and DDMCG
solutions depend on the number of neighbor nodes $|\mathcal{N}_k|$
of node $k$. The parameter $M$ is the length of the unknown vector
$\boldsymbol\omega_0$ that needs to be estimated.
\begin{table}
\centering
\caption{Computational Complexity Of Different Diffusion Algorithms}
\begin{tabular}{|c|c|c|}
\hline
Algorithm&Additions&Multiplications\\
\hline
DDCCG&$M^2+M$&$2M^2+2M$\\
&$+J(M^2+6M$&$+J(M^2+7M$\\
&$+|\mathcal{N}_k|M-4)$&$+|\mathcal{N}_k|M+3)$\\
\hline
DDMCG&$2M^2+10M-4$&$3M^2+12M+3$\\
&$+|\mathcal{N}_k|M$&$+|\mathcal{N}_k|M$\\
\hline
Diffusion LMS \cite{Cattivelli3}&$4M-1+|\mathcal{N}_k|M$&$3M+1+|\mathcal{N}_k|M$\\
\hline
Diffusion RLS \cite{Cattivelli2} &$4M^2+16M+1+|\mathcal{N}_k|M$&$4M^2+12M-1+|\mathcal{N}_k|M$\\
\hline
\end{tabular}
\label{table3:cc2}
\end{table}

\section{Preconditioner Design}

Preconditioning is an important technique which can be used to
improve the performance of CG algorithms
\cite{Benzi,Eisenstat,Knyazev,Axelsson1}. The idea behind
preconditioning is to employ the CG algorithms on an equivalent
system or in a transform--domain. Thus, instead of solving
$\boldsymbol R\boldsymbol\omega=\boldsymbol b$ we solve a related
problem $\tilde{\boldsymbol
R}\tilde{\boldsymbol\omega}=\tilde{\boldsymbol b}$, which is
modified with the aim of obtaining better convergence and steady
state performances. The relationships between these two equations
are given by
\begin{equation}
\tilde{\boldsymbol R} = \boldsymbol T\boldsymbol R\boldsymbol T^H,
\end{equation}
\begin{equation}
\tilde{\boldsymbol\omega} = \boldsymbol T\boldsymbol\omega
\end{equation}
and
\begin{equation}
\tilde{\boldsymbol b} = \boldsymbol T\boldsymbol b,
\end{equation}
where the $M\times M$ matrix $\boldsymbol T$ is called a preconditioner. We design the matrix $\boldsymbol T$ as an arbitrary unitary matrix of size $M\times M$ and has the following property \cite{Sayed}
\begin{equation}
\boldsymbol T\boldsymbol T^H = \boldsymbol T^H\boldsymbol T=\boldsymbol I.
\end{equation}

Two kinds of unitary transformations are considered to build the
preconditioner $\boldsymbol T$, which are discrete Fourier transform
(DFT) and discrete cosine transform (DCT) \cite{Sayed}. The
motivation behind employing these two matrix is they have useful
de--correlation properties and often reduce the eigenvalue spread of
the auto--correlation matrix of the input signal \cite{Sayed}.

For the DFT scheme, we employ the following expression
\begin{equation}
[\boldsymbol T_{DFT}]_{vm}\triangleq\frac{1}{\sqrt{M}}e^{-\frac{j2\pi mv}{M}},~~~ v,m=0,1,2,\ldots, M-1,
\end{equation}
where $v$ indicates the row index and $m$ the column index. $M$ is the length of the unknown parameter $\boldsymbol \omega_0$. The matrix form of $\boldsymbol T_{DFT}$ is illustrated as
\begin{equation}
\boldsymbol T_{DFT}= \frac{1}{\sqrt{M}}\left[
 \begin{matrix}
   1 & 1 & 1 & \cdots & 1\\
   1 & e^{-\frac{j2\pi}{M}} & e^{-\frac{j4\pi}{M}} & \cdots & e^{-\frac{j2(M-1)\pi}{M}}\\
   1 & e^{-\frac{j4\pi}{M}} & e^{-\frac{j8\pi}{M}} & \cdots & e^{-\frac{j4(M-1)\pi}{M}}\\
   \vdots & \vdots & \vdots & \ddots & \vdots\\
   1 & e^{-\frac{j2(M-1)\pi}{M}} & e^{-\frac{j4(M-1)\pi}{M}} & \cdots & e^{-\frac{j2(M-1)^2\pi}{M}}
  \end{matrix}
  \right]
\end{equation}

For the DCT scheme, the preconditioner
$\boldsymbol T$ is defined as
\begin{equation}
[\boldsymbol T_{DCT}]_{vm}\triangleq\delta(v)\cos\bigg(\frac{v(2m+1)\pi}{2M}\bigg),~~~ v,m=0,1,2,\ldots, M-1,
\end{equation}
where
\begin{equation}
\delta(0)=\frac{1}{\sqrt{M}}\ \ \textrm{and} \ \ \delta(v)=\sqrt{\frac{2}{M}}\  \textrm{for}\ v\neq 0
\end{equation}
and the matrix form of $\boldsymbol T_{DCT}$ is illustrated as
\begin{equation}
\boldsymbol T_{DCT}= \frac{1}{\sqrt{M}}\left[
 \begin{matrix}
   1 & 1 & 1 & \cdots & 1\\
   1 & \sqrt{2}\cos(\frac{3\pi}{2M}) & \sqrt{2}\cos(\frac{5\pi}{2M}) & \cdots & \sqrt{2}\cos(\frac{(2M-1)\pi}{2M})\\
   1 & \sqrt{2}\cos(\frac{6\pi}{2M}) & \sqrt{2}\cos(\frac{10\pi}{2M}) & \cdots & \sqrt{2}\cos(\frac{2(2M-1)\pi}{2M})\\
   \vdots & \vdots & \vdots & \ddots & \vdots\\
   1 & \sqrt{2}\cos(\frac{3(M-1)\pi}{2M}) & \sqrt{2}\cos(\frac{5(M-1)\pi}{2M}) & \cdots & \sqrt{2}\cos(\frac{(2M-1)(M-1)\pi}{2M})
  \end{matrix}
  \right]
\end{equation}
Then, for the DCT scheme, we choose $\boldsymbol T=\boldsymbol T_{DCT}^H$. It should be noticed that the scaling factor $\frac{1}{\sqrt{M}}$ is added in the expression for the $\boldsymbol T_{DFT}$ in order to result in a unitary transformation since then $\boldsymbol T_{DFT}$ satisfies $\boldsymbol T_{DFT}\boldsymbol T_{DFT}^H=\boldsymbol T_{DFT}^H\boldsymbol T_{DFT}=\boldsymbol I$ \cite{Sayed}.

The optimal selection of the preconditioner is the Kahunen--Lo$\grave{\textrm{e}}$ve transform (KLT) \cite{Sayed}. However, using KLT is not practical since it requires knowledge of the auto--correlation matrix $\boldsymbol R$ of the input signal and this information is generally lacking in implementations.

\section{Simulation Results}

In this section, we investigate the performance of the proposed
incremental and diffusion distributed CG--based algorithms in two
scenarios: distributed estimation and distributed spectrum
estimation in wireless sensor networks.

\subsection{Distributed Estimation in Wireless Sensor Networks}
In this subsection, we compare the proposed incremental and diffusion distributed CG--based algorithms with LMS \cite{Lopes1,Cattivelli3} and RLS \cite{Lopes1,Cattivelli2} algorithms, based on the MSE and MSD performance metrics. For each comparison, the number of
time instants is set to 1000, and we assume there are 20 nodes in the
network. The length of the unknown parameter $\boldsymbol \omega_0$ is 10, the
variance for the input signal and the noise are 1 and 0.001,
respectively. In addition, the noise samples are modeled as circular
Gaussian noise with zero mean.

\subsubsection{Performance of Proposed Incremental Distributed CG--Based Algorithms}
First, we define the parameters of the performance test for
each algorithm and the network. The step size $\mu$ for the LMS algorithm \cite{Lopes1} is set to 0.005, the forgetting factor $\lambda$ for the RLS \cite{Lopes1} algorithm is set to 0.998. The $\lambda_f$ for IDCCG and IDMCG are both set to 0.998. For IDMCG, the $\eta_f$ is equal to 0.55. The iteration number $J$ for IDCCG is set to 5. We choose the DCT matrix as the preconditioner.

\begin{figure}[!htb]
\begin{center}
\def\epsfsize#1#2{0.65\columnwidth}
\epsfbox{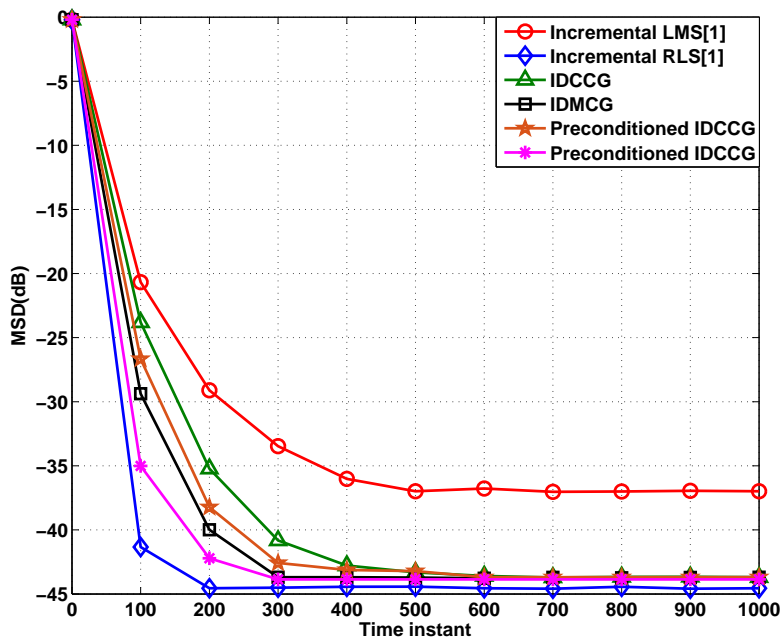}\vspace{-2.65em}\caption{\footnotesize MSD
performance comparison for the incremental distributed
strategies}\label{fig3:IDCG_simulation}
\end{center}
\end{figure}

\begin{figure}[!htb]
\begin{center}
\def\epsfsize#1#2{0.65\columnwidth}
\epsfbox{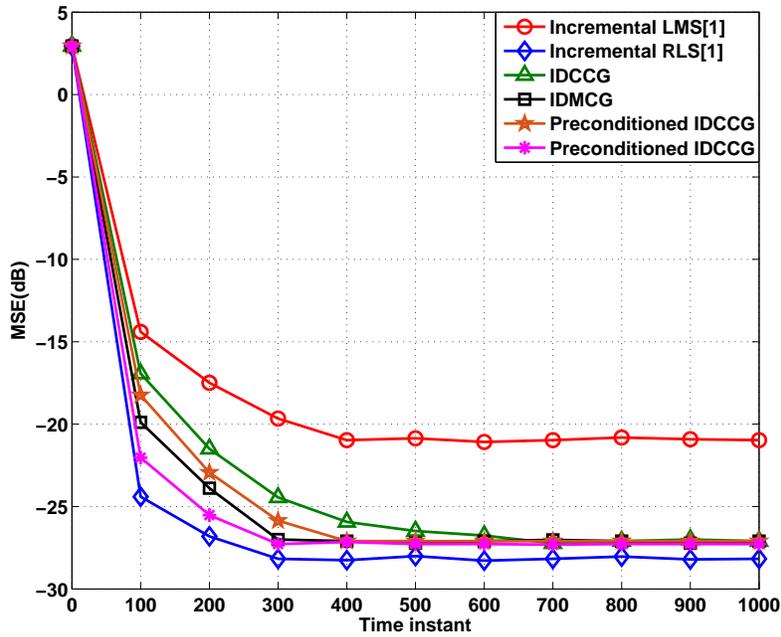}\vspace{-2.65em} \caption{\footnotesize MSE
performance comparison for the incremental distributed
strategies}\label{fig3:IDCG_simulation_MSE}
\end{center}
\end{figure}

The MSD and MSE performances of each algorithm have been shown in Fig. \ref{fig3:IDCG_simulation} and \ref{fig3:IDCG_simulation_MSE} respectively. We can verify that, the IDMCG and IDCCG algorithm performs
better than incremental LMS, while IDMCG is close to the RLS algorithm. With the preconditioning strategy, the performance of the IDCCG and IDMCG is further improved. The reason why the proposed IDMCG algorithm has a
better performance than IDCCG is because IDMCG employs the negative gradient vector
$\boldsymbol g_{k}$ with a recursive expression and the $\beta_{k}$
is computed using the Polak--Ribiere approach, which results in more accurate estimates. Comparing with the
IDCCG algorithm, the IDMCG is a non--reset and low complexity
algorithm with one iteration per time instant. Since the frequency
which the algorithm resets influences the performance, the IDMCG
algorithm introduces the non--reset method together with the Polak--
Ribiere approach which are used to improve the performance
\cite{Chang}.


\subsubsection{Performance of Proposed Diffusion Distributed CG--Based Algorithms}

The parameters of the performance test for each algorithm and the
network are defined as follows: the step size $\mu$ for the LMS
\cite{Cattivelli3} algorithm is set to 0.045, the forgetting factor
$\lambda$ for the RLS \cite{Cattivelli2} algorithm is set to 0.998.
The $\lambda_f$ for DDCCG and DDMCG are both 0.998. The $\eta_f$ is
equal to 0.45 for DDMCG. The iteration number $J$ for DDCCG is set
to 5. We choose the DCT matrix as the preconditioner.

\begin{figure}[!htb]
\begin{center}
\def\epsfsize#1#2{0.65\columnwidth}
\epsfbox{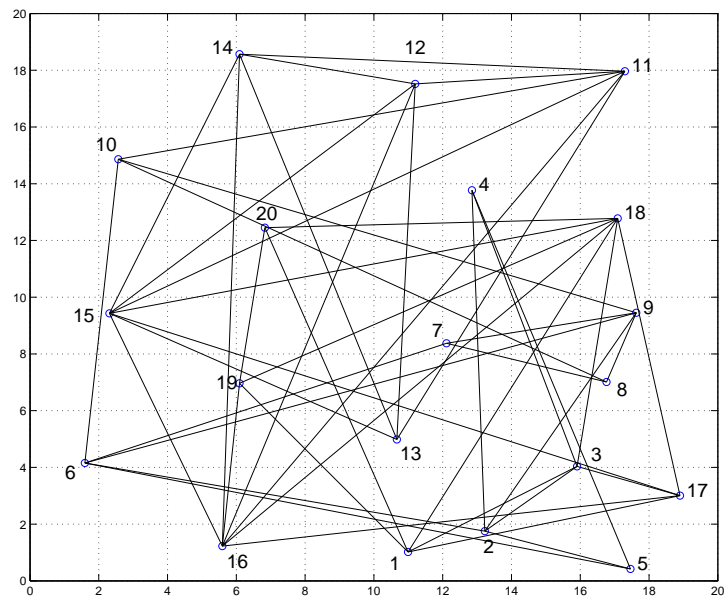} \vspace{-2.65em} \caption{\footnotesize Network
structure}\label{fig3:network}
\end{center}
\end{figure}

\begin{figure}[!htb]
\begin{center}
\def\epsfsize#1#2{0.65\columnwidth}
\epsfbox{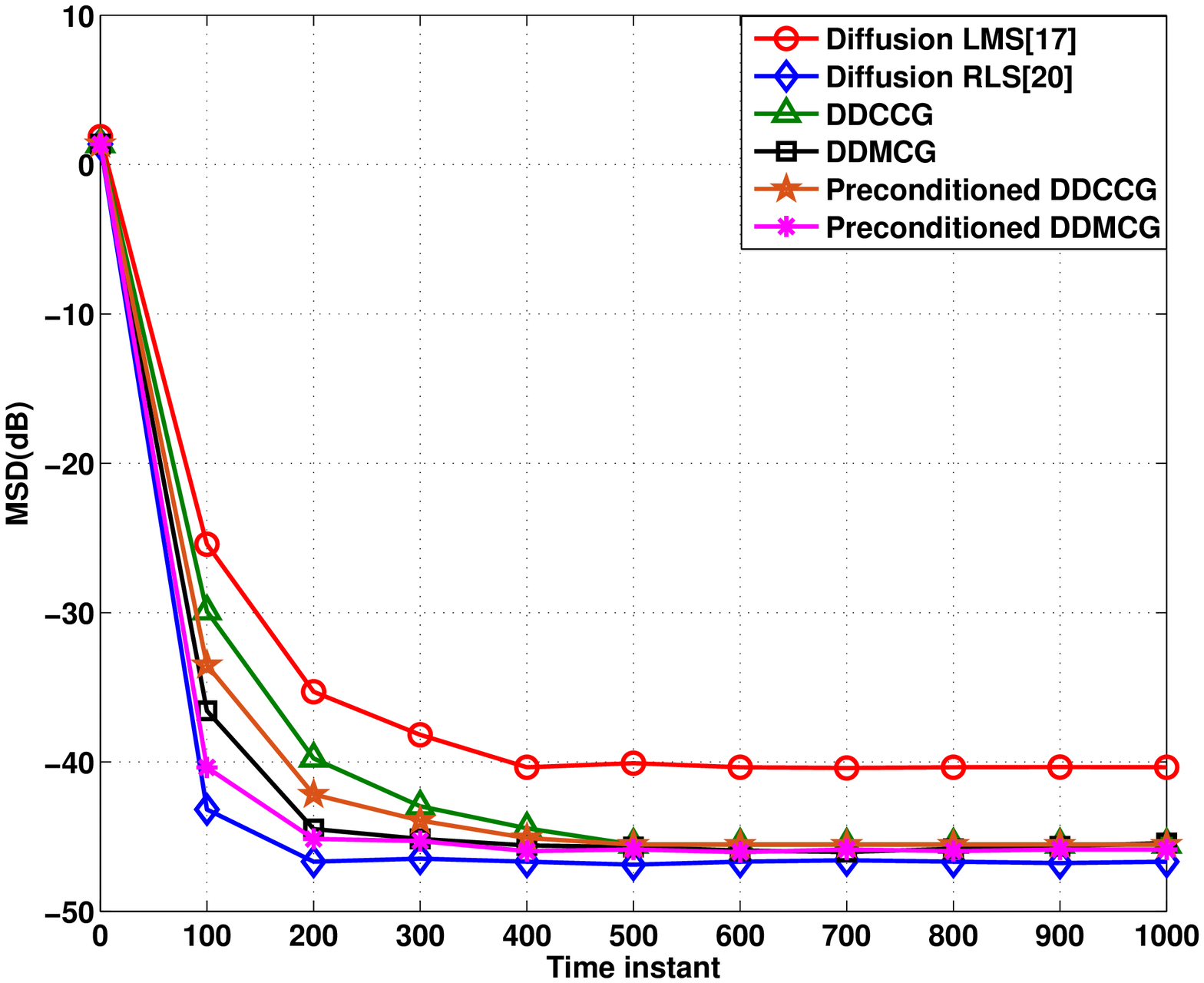} \vspace{-2.65em}\caption{\footnotesize MSD
performance comparison for the diffusion distributed strategies}
\label{fig3:DDCG_simulation}
\end{center}
\end{figure}

\begin{figure}[!htb]
\begin{center}
\def\epsfsize#1#2{0.65\columnwidth}
\epsfbox{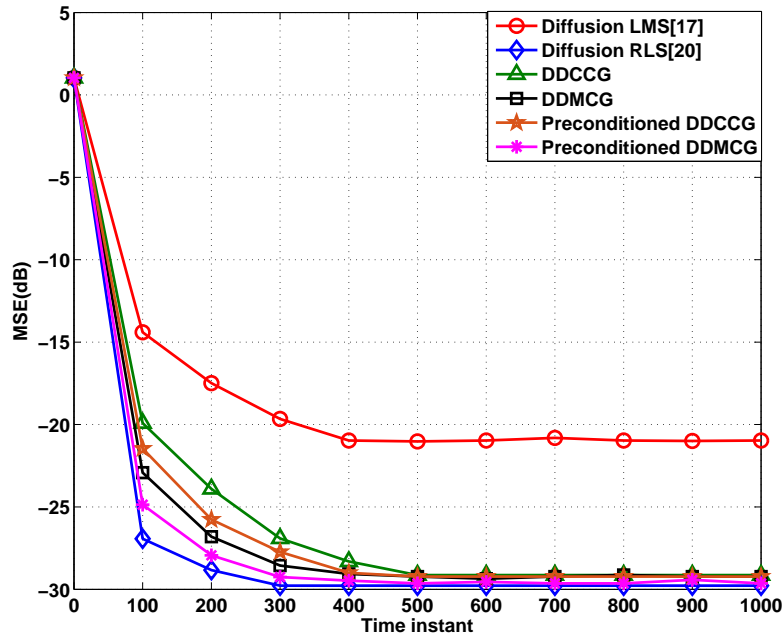}\vspace{-2.65em} \caption{\footnotesize MSE
performance comparison for the diffusion distributed strategies}
\label{fig3:DDCG_simulation_MSE}
\end{center}
\end{figure}

For the diffusion strategy, the combine coefficients $c_{kl}$ are calculated following the Metropolis rule. Fig. \ref{fig3:network} shows the network structure. The
results are illustrated in Fig. \ref{fig3:DDCG_simulation} and \ref{fig3:DDCG_simulation_MSE}. We can see that, the proposed
DDMCG and DDCCG still have a better performance than the LMS
algorithm and DDMCG is closer to the RLS's performance. The performance of the DDCCG and DDMCG can still benefit from the preconditioning strategy.


\subsection{Distributed Spectrum Estimation}
In this simulation, we consider a network composed of $N=20$ nodes estimating the unknown spectrum $\boldsymbol\omega_0$, as illustrated in Fig. \ref{fig3:network}. The nodes scan $N_c=100$ frequencies over the frequency axis, which is normalized between 0 and 1, and use $\mathcal{B}=50$ non--overlapping rectangular basis functions to model the expansion of the spectrum \cite{Lorenzo2}. The basis functions have amplitude equal to one. We assume that the unknown spectrum $\boldsymbol\omega_0$ is transmitted over 8 basis functions, thus leading to a sparsity ratio equal to 8/50. The power transmitted over each basis function is set equal to 1.

For distributed estimation, we employ the DDMCG and the DDCCG algorithms, together with the preconditioned DDMCG algorithm to solve the cost function (\ref{Eqn3:cf_se2}) respectively. The $\lambda_f$ for DDCCG and DDMCG are both 0.99. The $\eta_f$ is equal to 0.3 for DDMCG. The iteration number $J$ for DDCCG is set to 5. The DCT matrix is employed as the preconditioner. We compare the proposed DDCCG and DDMCG algorithms with the sparse ATC diffusion algorithm \cite{Lorenzo2}, diffusion LMS algorithm \cite{Cattivelli3} and diffusion RLS algorithm \cite{Cattivelli2}. The step--sizes for the sparse ATC diffusion algorithm and diffusion LMS algorithm are set equal to 0.05, while for the sparse ATC diffusion algorithm, $\gamma$ is set to $2.2\times 10^-3$ and $\beta$ is set to 50. The forgetting factor $\lambda$ for the diffusion RLS algorithm is set to 0.998.

\begin{figure}[!htb]
\begin{center}
\def\epsfsize#1#2{0.65\columnwidth}
\epsfbox{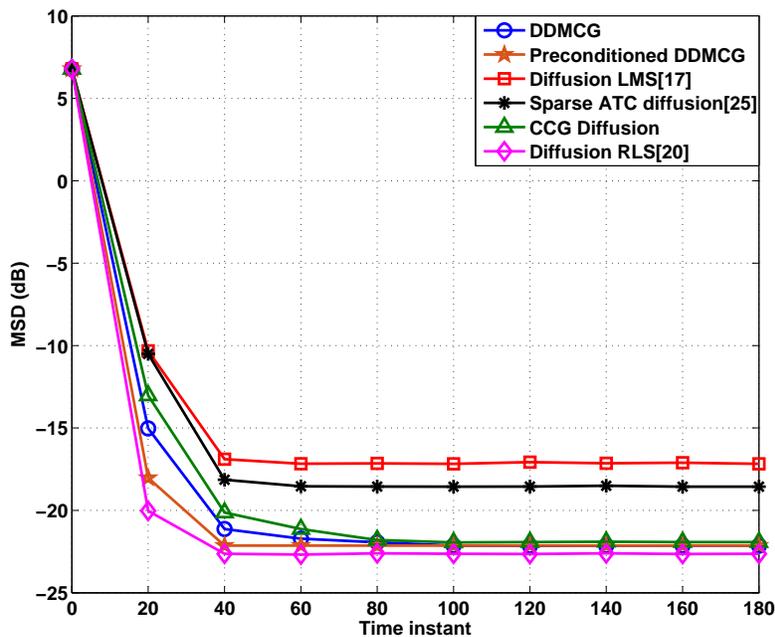} \vspace{-2.65em}\caption{\footnotesize
Performance comparison for the distributed spectrum estimation}
\label{fig3:spectrum_estimation}
\end{center}
\end{figure}

\begin{figure}[!htb]
\begin{center}
\def\epsfsize#1#2{0.65\columnwidth}
\epsfbox{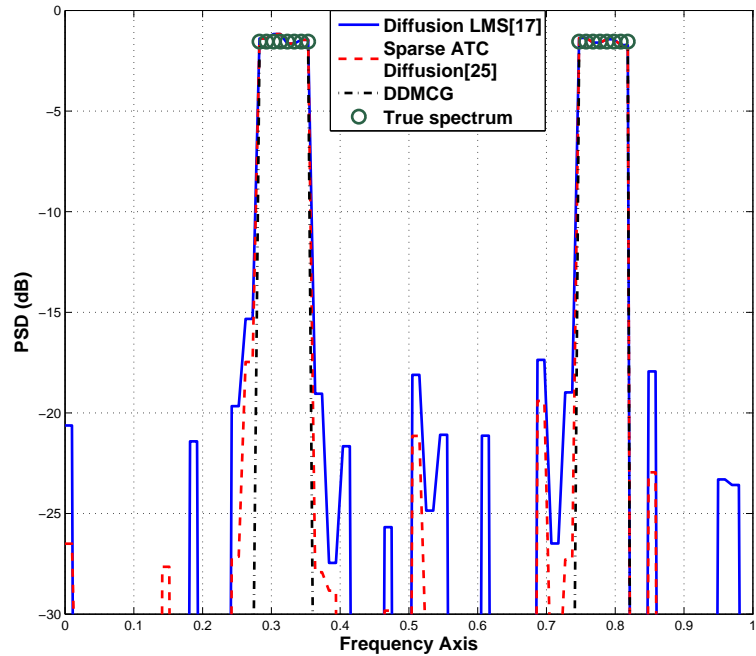} \vspace{-2.65em}\caption{\footnotesize Example of
distributed spectrum estimation} \label{fig3:PSD}
\end{center}
\end{figure}

We illustrate the result of distributed spectrum estimation carried out by different algorithms in the term of the MSD comparison in Fig. \ref{fig3:spectrum_estimation}. We also select the sparse ATC diffusion algorithm \cite{Lorenzo2}, diffusion LMS algorithm \cite{Cattivelli3} and DDMCG to compare their performance in term of PSD in Fig. \ref{fig3:PSD}. The true transmitted spectrum is also reported in Fig. \ref{fig3:PSD}.

From Fig. \ref{fig3:spectrum_estimation}, the DDMCG still performs
better than other algorithms and is close to the diffusion RLS
algorithm. From Fig. \ref{fig3:PSD}, we can notice that all the
algorithms are able to identify the spectrum, but it is also clear
that the DDMCG algorithm is able to strongly reduce the effect of
the spurious terms.

\section{Conclusions}

In this paper, we have proposed distributed CG algorithms for both
incremental and diffusion adaptive strategies. We have investigated
the proposed algorithms in distributed estimation for wireless
sensor networks and distributed spectrum estimation. The CG--based
strategies has low computational complexity when compared with the
RLS algorithm and have a faster convergence than the LMS algorithm.
The preconditioning strategy is also introduced to further improve
the performance of the proposed algorithms. Simulation results have
proved the advantages of the proposed IDCCG/IDMCG and DDCCG/DDMCG
algorithms in different applications.
\bibliographystyle{IEEEtran}
\bibliography{reference}

\end{document}